\documentclass{emulateapj}

\usepackage[dvips]{color}

\newcommand{\mt}[1]{\mathrm{#1}}
\def\kms{\textrm{ km s}^{-1}}

\shortauthors{HAYASHI \& CHIBA}
\shorttitle{Size Evolution of a Galactic Disk}

\begin{document}

\title{On the Size Evolution of a Galactic Disk
in Hierarchical Merging of Cold Dark Matter Halos}

\author{Hirohito~Hayashi and Masashi~Chiba}

\affil{Astronomical Institute, Tohoku University,
Aoba-ku, Sendai 980-8578, Japan}

\begin{abstract}
We investigate the dynamical effects of dark matter subhalos on the structure
and evolution of a galactic disk, using semi-analytic method that includes 
approximated and empirical relations as achieved in detailed numerical
simulations of the cold dark matter model. We calculate the upper limit for
the size of a galactic disk at a specific redshift $z$, based on the orbital
properties of subhalos characterized by their pericentric distances from the center
of a host halo. We find that this possibly largest size of a disk as determined
by the smallest pericentric distances of subhalos shows the characteristic
properties, which are basically in agreement with an observed galactic disk
at low and high $z$. Namely, it is found that a massive disk can have
a larger size than a less massive one, because of its stability against
the destruction effect of subhalos. Also, with fixed mass, the size of
a galactic disk at low $z$ can be larger than that at high $z$, reflecting
the orbital evolution of subhalos with respect to a host halo.
These results suggest that the presence and structure of a galactic disk
may be dynamically limited by the interaction with dark matter substructures,
especially at high $z$.
\end{abstract}


\section{INTRODUCTION}

The cold dark matter (CDM) model is now a standard theory for the 
structure formation of the universe. The model provides us with a basic 
theoretical framework for understanding the hierarchical clustering of 
dark matter, where large dark halos are assembled via merging and 
accretion of smaller halos. Indeed, the prediction of the CDM model is
in excellent agreement with observations of large-scale structures
(Tegmark et al. 2004). However, recent high-resolution N-body simulations 
based on the CDM model have highlighted that as a result of these merging
processes, a large dark halo like that of the Milky Way contains numerous
dark matter substructures or subhalos inside its virial radius,
in contrast to the small number of known Milky Way satellites
(Klypin et al. 1999; Moore et al. 1999). This so-called missing
satellite problem remains unsolved yet, although various ideas are proposed
to solve it based on, e.g., astrophysical baryonic processes such as
UV feedback to suppress star formation and/or observational selection effects
for faint satellite galaxies in the Milky Way
(e.g., Bullock, Kravtsov \& Weinberg 2000; Madau et al. 2008;
Tollerud et al. 2008; Koposov et al. 2008; Koposov et al. 2009;
Macci\`o et al. 2009).

If such dark matter subhalos indeed exist around a disk galaxy and some of these,
especially massive ones, interact with a galactic disk in the course of their 
orbital motions, then the stellar component of a galactic disk can be made 
so thick due to dynamical heating that it no longer exists as a thin stellar disk
like that of the Milky Way at the present epoch. This issue has been investigated
by many researchers as a constraint on models of galaxy formation and evolution
(e.g., T\'oth \& Ostriker 1992; Font et al. 2001; Ardi et al. 2003;
Hayashi \& Chiba 2006; Villalobos \& Helmi 2008; Hopkins et al. 2008;
Purcell et al. 2009). Hopkins et al. (2008) recently showed, based on the realistic
(radial) orbits of subhalos as obtained from cosmological simulations, that
disk heating is less powerful than previously thought by T\'oth \& Ostriker (1992),
which is based on the simple assumption of rigid satellites with circular orbits.
To test what this result implies for satellite accretion with high mass,
Purcell et al. (2009) made high-resolution numerical simulations for
satellite-disk interactions and showed that in reality
accretion events of mass ratio $\sim$1:10 do not preserve thin disk components.
Indeed, Stewart et al. (2008) demonstrated, also based on N-body 
simulations, that a majority ($\sim 70 \% $) of galaxy-sized halos 
with mass $M=10^{12}M_\odot$ at $z = 0$ have accreted at least one object 
with $M > 10^{11}M_\odot$ over the last 10 Gyr and $\sim 95\%$ have 
accreted an object with mass greater than Milky Way disk ($M > 5\times 
10^{10}M_\odot$). As these works suggest, a galactic disk is a fragile system,
so that its dynamical evolution may be severely limited by these subhalos. In 
particular, the presence of a thin galactic disk at the present epoch 
implies the absence of the effect of subhalos within its size, which 
suggests in turn that the size of a galactic disk may be regulated by 
the orbital radii of subhalos. Therefore it is important to investigate 
the relation between the size evolution of a galactic disk and dynamics 
of subhalos within a host halo.

Several models for the structural evolution of a galactic disk have been studied
(Mo, Mao \& White 1998, hereafter MMW98; Bouwens \& Silk 2002, hereafter
BS02; Kampakoglou \& Silk 2007). The evolution
models can be divided into two alternative approaches, one is forward approach
and the other is backward approach.
The forward approach is based on the CDM model, where both collapsing dark
matter and baryonic gas acquire the angular momentum through tidal force
and mergers. When a gas component cools, condenses and forms stars, then the system's
angular momentum halts the collapse, leading to the formation of a rotationally
supported disk (Fall \& Efstathiou 1980). Under the assumption that the
fractions of disk mass and angular momentum in the disk relative to the
halo, together with the spin parameter of the halo, do not vary during this
collapsing phase, Mo, Mao \& White (1999) obtained the following relations for
the scale length of an exponential stellar disk:
$R_d \propto H(z)^{-1}$ at a fixed circular velocity or
$R_d\propto H(z)^{-2/3}$ at a fixed halo mass, where $H(z)$ is a Hubble parameter
at $z$. On the other hand, the
backward approach is based on the detailed models of the properties of
local disk galaxies, e.g., gaseous, stellar, and metallicity profiles, as well as
current star formation rate (SFR) and age-metallicity
relationship. This approach uses the local universe as a reference. Then
the properties of a galactic disk at high $z$ are derived on the
basis of those at present time and backward calculation in time. Using the Milky
Way as reference and assuming the continuous infall of metal-free gas
from outside, BS02 provide the following size-redshift relation for a disk:
$r(z)/r(0)=1-0.27z$.

Observational studies of a galactic disk have also been put forward by 
several researchers. Using the Sloan Digital Sky Survey (SDSS), Shen et al. 
(2003) showed the local size-mass relation, where a galactic disk with 
large mass has a larger scale length than that with small mass. Based on 
the Galaxy Evolution from Morphology and SEDs (GEMS) survey, Barden et 
al. (2005) presented the size-mass relation up to $z \sim 1$. Also, 
using the Faint Infrared Extragalactic Survey (FIRES), Trujillo et al. (2004) 
measured the size-mass relation in the rest-frame optical up to $z\sim 2.5$. 
Trujillo et al. (2006) presented that at a given mass the 
mean size of a galactic disk was $\sim 2$ times smaller at $z \sim 
2.5$ than that we see today, using the result of FIRES, GEMS, and SDSS.

It is worth noting that models in both forward and backward approaches are
in good agreement with these observational results.
However, there are a couple of unresolved
issues in these models. In forward approach, the model
assumes that a gas disk forms a stellar disk instantaneously at the epoch
of a halo virialization, whereby only the evolution of a host halo is taken
into account. In backward approach, the model does not include the merging
history of a host halo. Thus in these models, the important dynamical effects
of subhalos on the structural evolution of a galactic disk are unclear.
In particular, since subhalos are able to destroy a galactic disk and the
probability of such events is expected to be high, the previous models
are yet insufficient to understand the evolution of a galactic disk.
Therefore, it is of great importance to fully take into account the effect
of subhalos on the evolution model of a galactic disk.

We remark here that this aspect of a disk evolution is automatically included in
high-resolution gasdynamical simulations of a collapsing galaxy in the
context of the CDM model. However, many of such simulation works have produced
a disk which is too small compared with a galactic disk at present time
(e.g., Navarro, Frenk \& White 1995), most probably because the simulation
results are largely affected by yet poorly understood physical processes,
such as star formation and supernova feedback, that work below
the resolution limit.
Indeed, Governato et al. (2007) showed that their simulation taking into account
improved feedback models along with high numerical resolution largely minimizes
the issue of a too small simulated disk.
Here, to avoid such complexity in understanding the calculated results,
we concentrate on the dynamical effect of subhalos
on a galactic disk and adopt a semi-analytic method that includes approximated
and empirical relations as achieved in detailed numerical simulations of the
CDM model.

The outline of this paper is as follows. In \S\ref{model} we
describe our semi-analytical model that we use for the merging and accretion
of dark matter halos. In \S\ref{results}
we present the examples of our semi-analytical model and the results for
the disk evolution. We summarize and discuss the implications of the
current results in \S\ref{discussion}.

\section{MODEL}\label{model}

We describe here the properties of subhalos in a host dark halo, 
in particular the mass accretion history and the orbital evolution of 
subhalos after they enter the virial radius of a host halo.
For our current work, we adopt the semi-analytic
model developed by Zentner \& Bullock (2003), hereafter ZB03, with some
modification as explained below. In the followings, we adopt the standard set of 
cosmological parameters: the density parameter at present 
$\Omega_{\mt{m0}}=0.3$, cosmological constant at present $\Lambda=0.7$, 
Hubble constant $H_0=100h \kms \mathrm{Mpc^{-1}}$ with 
$h = 0.72$, baryon fraction at present $\Omega_{\mt{B}} h^2 = 0.02$, and
rms density variation averaged over $8h^{-1}$ Mpc as $\sigma_8=0.95$.

\subsection{The ZB03 Model}

\subsubsection{Construction of Merger Trees}

Following ZB03, we track the mass accretion history of a host halo using the 
extended Press-Schechter (EPS) theory (Bond et al. 1991; 
Lacey \& Cole 1993, hereafter LC93) and adopt the merger tree algorithm 
proposed by Somerville \& Kollat (1999, hereafter SK99).
Using this method, we generate a list of the masses and accretion redshifts
of all subhalos that have merged to form a host halo of a given mass
at a given redshift. In the EPS formalism, the linear density field
$\delta(M)$ as a function of the smoothing mass $M$ is regarded as a random
Gaussian field, which is specified by the mass variance $S(M)$. To derive this,
we consider a density field smoothed with a spherical top-hat function
with a radius $R$. Also, according to the spherical collapse model, 
the density field collapses and forms a virialized object at $z$ 
once the density field $\delta$ exceeds $\omega 
=\delta_{\mt{crit}}(z)\simeq$ 1.68 (see LC93).
Then, denoting $\Delta S = S(M) - S(M + \Delta M)$ and
$\delta \omega = \omega(t) - \omega(t + \Delta t)$,
the probability that a halo having mass $M$ at time $t$ has
accreted the mass corresponding to a step of $\Delta S$
in a time step associated with $\delta \omega$ is given as
\begin{equation}
 P(\Delta S,\delta \omega)d(\Delta S)=\frac{\delta \omega}{\sqrt{2\pi}\Delta
 S^{3/2}}\exp{\left[-\frac{(\delta \omega)^2}{2\Delta S}\right]}d(\Delta S).
 \label{EPS-prob}
\end{equation}

Following SK99, we generate merger trees by starting the redshift $z=0$ at
which we consider the final halo and step backward in time.
To reproduce the prediction of the conditional mass 
function of the EPS model, we must choose the appropriate time step. 
In this model we choose the time step given by Taylor \& Babul (2004),
$\delta \omega = [0.2\log_{10}{ (M/M_{\mt{min}})}+0.1]\delta \omega_0$,
where $M_{\mt{min}}$ is a resolution limit mass and $\omega_0 =
 \sqrt{(dS/dM)|_{M}M_{\mt{min}}}$.
At each timestep we take a mass $S(M_\mt{p})=S(M) + \Delta S$ from the 
probability equation (\ref{EPS-prob}). We treat events with $M_{\mt{p}} < 
M_{\mt{min}}$ as accreted mass and retain all information about mergers 
with $M_{\mt{p}} \ge M_{\mt{min}}$. In this way, we generate the list of 
progenitor masses and accretion redshifts at each time step. This 
process continues until the  masses of all progenitors are less than 
$M_{\mt{min}}$. At each time step we identify the most massive 
progenitor with the host halo and all less massive progenitor with 
accreted subhalos. 

We note here that the recent work by Zhang, Fakhouri \& Ma (2008) showed
cautionary remarks in the use of the SK99 model, which overestimates the
abundances of small progenitor halos as large as a factor of about 2
compared to the prediction of the EPS formalism. Although a quantitative
estimation for the effect of adopting different Monte Carlo algorithms
from SK99 is beyond the scope of this work, our results based on
the distribution of pericentric orbital radii of subhalos,
as detailed later, would not be sensitive to their precise abundances
at each redshift.

\subsubsection{Density Distribution of Halos}

The spherical collapse model provides the density of a virialized region 
for a dark halo. The density of a virialized halo is given as 
$\rho_{\mt{vir}}=\rho_{\mt{m}} \Delta_{\mt{vir}}$, where
$\rho_{\mt{m}}$ is the mean matter density of the universe.
For $\Delta_{\mt{vir}}$, we use the fitting function provided by
Bryan \& Norman (1998), where for our cosmological parameters,
$\Delta_{\mt{vir}}\approx 370$ at $z = 0$ and
$\Delta_{\mt{vir}}\approx 179$ at high $z$.
Then the virial radius of a virialized halo with mass $M_{\mt{vir}}$ at $z$ is 
\begin{equation}
R_{\mt{vir}} = (3M_{\mt{vir}} / 4\pi \Delta_{\mt{vir}} \rho_{\mt{m}})^{1/3}
 \label{virial-radius}
\end{equation}
and the circular velocity at the virial radius, so-called virial velocity, is
$V_{\mt{vir}}=(GM_{\mt{vir}}/R_{\mt{vir}})^{1/2}$.

Recent numerical simulations discover the several analytical density 
profile of a dark halo (Navarro, Frenk \& White 1997, hereafter NFW; 
Moore et al. 1998). We choose the NFW profile as the density profile of 
all halos:
\begin{equation}
 \rho(r)=\frac{\rho_s}{(r/r_s)(1+r/r_s)^2} \ ,
\end{equation}
where the maximum circular velocity is achieved at
$r_{\mt{max}} \simeq 2.16r_s$.
The NFW profile is
charaterized by the concentration parameter
$c_{\mt{vir}} =R_{\mt{vir}}/r_s$. Numerical simulations 
suggest several analytical functions with the mass of a halo and a
redshift, $c_{\mt{vir}}=c_{\mt{vir}}(M, z)$ (NFW; Bullock et al. 
2001, hereafter B01; Eke, Navarro \& Steinmetz 2001; Macci\`o, Dutton \& 
van den Bosch 2008). In our model, we use the model of B01
for the concentration parameter of a dark matter halo having mass 
$M_{\mt{vir}}$ at redshift $z$, which has collapsed at redshift $z_c$:
$c_{\mt{vir}}(M,z)=K (1+z_c)/(1+z)$,
where B01 suggest $K = 4$ since this agrees well with the result of their 
$\Lambda$CDM simulations.
This expression for $c_{\mt{vir}}(M,z)$ indicates that massive halos are 
less concentrated than less massive halos and that a halo at low $z$
is more concentrated than that at high $z$. Since 
$R_{\mt{vir}} \propto \Delta_{\mt{vir}}^{1/3}(1+z) $, an inner radius $r_s$ is
almost constant at all $z$.

\subsubsection{Orbital Evolution of Subhalos}

We track the orbital evolution of all accreted subhalos, taking into account the 
dynamical friction and tidal mass loss while orbiting in a host halo.
Through these processes, subhalos eventually sink to the center of 
the host halo (termed ``central merged'' in ZB03), or lose the most of the mass
(``tidal disrupted''), or keep the distinct structure from their accretion epoch to
the present. We model these processes based on the ZB03 formalism.

We define that a subhalo has a mass $M_{\mt{sat}}$ and outer radius
$R_{\mt{sat}}$ at the accretion redshift of $z_{\mt{acc}}$ obtained by
our merger tree, where $R_{\mt{sat}}$ is given by $R_{\mt{vir}}$ [equation 
(\ref{virial-radius})]. The concentration parameter of a subhalo is 
calculated from the median value as given by 
$M_{\mt{sat}}$ and $z_{\mt{acc}}$ at the accretion. Notice that 
$R_{\mt{sat}}$ and $M_{\mt{sat}}$ are varied due to mass loss while
orbiting within a host halo. We update the mass and concentration 
parameter of the host halo at each accretion event and continue to fix 
its density structure of the host halo while each orbit is integrated. 
In order to track the orbit of each subhalo, we assume the
potential of the host halo to be spherically symmetric and static,
so the orbit of a subhalo depends only on the radial distance
from the center of the host halo.

When it accretes onto the host halo, we assume that each subhalo has an 
initial orbital energy, which is consistent with the result of
detailed N-body simulations by, e.g., Klypin et al. (1999). Namely, as described
in ZB03, each subhalo has an initial orbital energy equal to the energy of 
a circular orbit at the radius $R_{\mt{circ}}=\eta R_{\mt{vir}}$, where 
$R_{\mt{vir}}$ is the virial radius at the accretion and $\eta$ is drawn 
randomly from a uniform distribution on the interval $[0.4, 0.75]$. In 
this way, we define the absolute velocity of a subhalo at the 
accretion. Next we consider the velocity vector of a subhalo. We 
define the initial angular momentum of a subhalo at the accretion as 
$J=\epsilon J_{\mt{circ}}$, where $J_{\mt{circ}}$ is the angular 
momentum for a circular orbit of the same energy and $\epsilon$ is a 
parameter, so-called ``orbital circularity'' (LC93). Recent numerical 
simulations show that this orbital circularity is well fitted by beta 
distribution (Zentner et al. 2005):
\begin{equation}
 \frac{df(\epsilon)}{d\epsilon}=\frac{\Gamma(2a)}{\Gamma^2(a)} \epsilon^{a-1} 
(1-\epsilon)^{a-1}, \label{eq-epsilon}
\end{equation}
where $a = 2.22$ provides a good fit to their $\Lambda$CDM simulation 
results with the mean $\left<\epsilon\right>=1/2$ and the dispersion 
$\sigma(\epsilon)=1/2(2a+1)^{1/2}$.
This form of the orbital circularity is also in agreement with the results
obtained from cosmological simulations by Benson (2005) and
Khochfar \& Burkert (2006).
We define that the initial radial position of a 
subhalo at the accretion is $R_{\mt{circ}}$ and its orbit 
is given by $dR/dt < 0$ in order to be initially infalling.
For the calculation of its orbital decay by dynamical friction, we assume
a subhalo as being represented by a point mass in the gravitational
potential of the host halo. We use the standard Chandrasekhar 
formula (Chandrasekhar 1943) for the evaluation of the frictional force,
following the method shown in ZB03.

The mass of a subhalo is stripped by the tidal force while orbiting in the 
potential of the host halo. This process is approximately modeled as follows.
First, the instantaneous tidal radius $r_t$ of the subhalo at each point along
its orbit is calculated by the equation, in the limit that a subhalo is much
smaller than a host halo, as (e.g., King 1962)
\begin{equation}
 r_t^3\simeq 
\frac{M_{\mt{sat}}(<r_t)/M_{\mt{host}}(<r)}{2+\omega^2r^3/GM_{\mt{host}}
(<r)-\partial \ln M_{\mt{host}}(<r)/\partial \ln r}r^3,
\end{equation}
where $r$ is the radial position of a subhalo in the host halo,
$M_{\mt{host}}(<r)$ is the mass of the host halo within $r$,
$M_{\mt{sat}}(<r_t)$ is the mass of the subhalo within $r_t$,
$\omega$ is the instantaneous angular speed of 
the orbiting subhalo $\omega = L / r^2$, and $L$ is the angular momentum 
of the subhalo. The process of the mass loss occuring outside $r_t$ is calculated
following Taylor \& Babul (2001): we divide the orbit of the subhalo
into discrete time steps of size $\delta t \ll T$, where $T$ is the orbital
time scale $T = 2 \pi / \omega$, then at each time step, we remove the mass
$\delta m = M_{\mt{sat}}(>r_t) (\delta t/T)$,
where $M_{\mt{sat}}(>r_t)$ is the mass of the subhalo outside $r_t$.
While a subhalo loses mass due to tidal force, its density profile is assumed to
remain unchanged within its outer radius $R_{\mt{sat}}$,
which is now defined as the radius within which the bound mass is contained
in the form of an NFW profile. We continue to fix the scale radius of
the subhalo, $r_s^{\mt{sat}}$, at the value calculated at the
epoch of the accretion.

Finally, we define whether a subhalo is tidally disrupted, centrally merged, or 
survived, following the criteria given in ZB03. We first set
$r_{\mt{max}}^{\mt{sat}}\simeq 2.16 r_s^{\mt{sat}}$ as the radius at which the 
initial circular-velocity profile of a subhalo reaches its maximum, and its mass
$M_{\mt{sat}}(<r_{\mt{max}}^{\mt{sat}})$ within $r_{\mt{max}}^{\mt{sat}}$.
We consider a subhalo to be ``tidally disrupted'' if its mass becomes less than 
$M_{\mt{sat}}(<r_{\mt{max}}^{\mt{sat}})$. A subhalo is ``centrally 
merged'' with the host halo if its radial position, $r$, becomes smaller
than $r_{\mt{max}}^{\mt{sat}}$. 
Also, the mass lost by tidal force modifies the circular-velocity profile
of subhalos. To account for this effect, following ZB03, we determine whether or not
the tidal radius of each surviving subhalo has ever been less than 
$r_{\mt{max}}^{\mt{sat}}$. If the subhalo experienced so, we change the maximum 
circular velocity of the subhalo by
$V_{\mt{max}}^{\mt{final}}= (M_{\mt{sat}}^{\mt{final}} /
  M_{\mt{sat}}^{\mt{initial}} )^{1/3} V_{\mt{max}}^{\mt{initial}}$,
where $V_{\mt{max}}^{\mt{initial}}$ is the maximum circular velocity of the 
initially defined subhalo, $M_{\mt{sat}}^{\mt{final}}$ is 
its final mass, and $M_{\mt{sat}}^{\mt{initial}}$ is its initial mass at the 
accretion.

\subsection{Model Checks}

\begin{figure}
\includegraphics[]{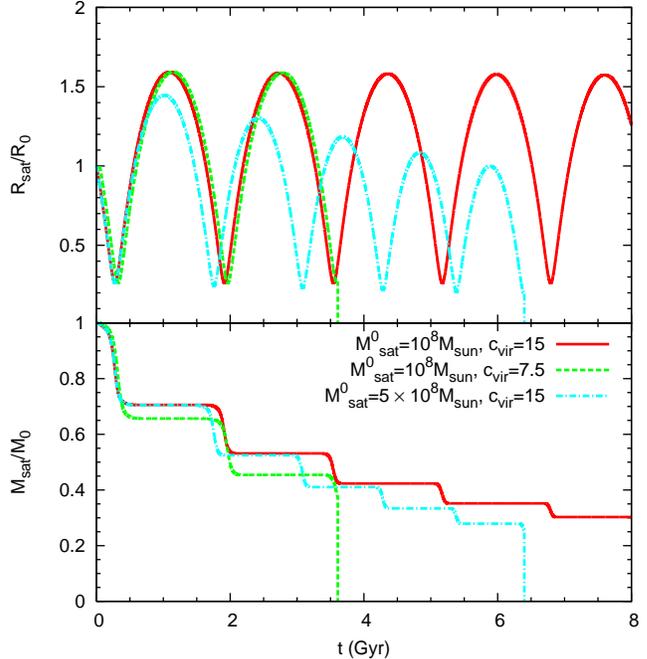}
\caption{Orbital 
evolution for three sets of the subhalo parameters: Initially at $t=0$, 
$M_{\mt{sat}}^0=10^8 M_\odot$, $c_{\mt{vir}} = 15$ (solid line); 
$M_{\mt{sat}}^0 = 10^8 M_\odot$, $c_{\mt{vir}}=7.5$ (dashed line); 
$M_{\mt{sat}}^0 = 5 \times 10^9 M_\odot$, $c_{\mt{vir}} = 15$
(dash-dotted line). The host halo is given by $c_{\mt{vir}}=6$ and
$M_{\mt{host}}=5\times 10^{11}M_\odot$. The top panel shows the radial evolution 
in units of the initial radius as a function of time. The bottom panel 
shows the mass evolution in units of the initial mass as a function of 
time, where the vertical lines correspond to the end point when the subhalos 
are destroyed.}
\label{fig-example1}
\end{figure}

\begin{figure}
\includegraphics[]{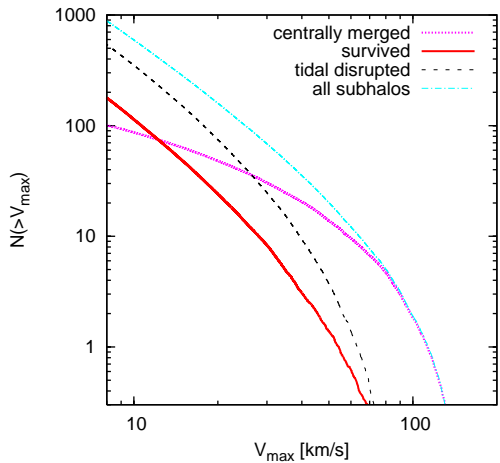}
\caption{Cumulative velocity 
function of the survived and destroyed subhalos for the host mass 
$M_{\mt{host}}=1.4\times 10^{14}M_\odot$ at $z = 0$. These plots are an 
ensemble mean of 200 host halo systems. Each line is for the survived 
subhalos until the present epoch (thick solid line), tidally destroyed 
ones (dashed line), central-merged ones (thick dotted line), and 
all of the accreted subhalos as obtained from the merging history 
(dash-dotted line), respectively.}
\label{fig-VF}
\end{figure}

Following ZB03, we have calculated the three examples for the orbits of
subhalos. For all of these cases, we adopt 
the same initial condition of $\epsilon = 0.5$ and $\eta=0.5$ for each 
subhalo's orbit, with the concentration and the mass of the host halo 
being $c_{\mt{vir}}=6$ and $M_{\mt{host}}=5\times 10^{11}M_\odot$, 
respectively. The other parameters of the subhalos are given as follows: 
$M_{\mt{sat}}^0=10^8M_\odot$, $c_{\mt{vir}}=15$; 
$M_{\mt{sat}}^0 = 10^8M_\odot$, $c_{\mt{vir}}=7.5$; 
$M_{\mt{sat}}^0 = 5\times 10^9M_\odot$, $c_{\mt{vir}} = 15$.
The results are shown in Figure \ref{fig-example1}, where
the upper panel shows the evolution of radial 
positions of the subhalos and the lower panel shows the evolution of 
their masses of the subhalos. We set the start of calculation to be at 8 
Gyr ago, which is $z \simeq 1.14$ for our cosmology.
The plots in this figure are in excellent agreement with Figure 3 in ZB03,
indicating that our model successfully reproduces their results.

We show, in Figure \ref{fig-VF}, the cumulative velocity function to 
demonstrate the statistical properties of our model. The plots correspond to an 
ensemble mean of 200 host halos with mass $1.4\times 10^{12}M_\odot$ at 
$z = 0$, where these host halo systems have the 
maximum circular velocity $V_{\mt{max}} \simeq 188 \kms$, the virial 
radius $R_{\mt{host}} \simeq 285$ kpc, and the concentration parameter 
$c_{\mt{vir}} \simeq 13.4$. We use $M_{\mt{min}}=10^{5} M_\odot$.
It is found that the cumulative velocity function of the 
survived subhalos until $z=0$ is in good agreement with Figure 4
of ZB03 and also the result of the numerical simulation by Klypin et al. (1999).

\subsection{Method for Calculating Likely Disk Sizes}

A host halo contains numerous subhalos, among which there exist several 
very massive ones with mass of $M \sim 10^{10}M_\odot$. In this study, we 
assume that when such a massive subhalo having a comparable mass to a 
galactic stellar disk of $M \sim 10^{10}M_\odot$ interacts with each 
other, then a stellar disk is made so thick due to dynamical heating 
that it no longer exists as a thin stellar disk like that of the Milky 
Way at the present epoch (e.g., Hayashi \& Chiba 2006; Villalobos \& Helmi 2008).
More specifically, we assume that if a subhalo having this large mass interacts
with a galactic disk at any radial positions, then the latter is supposed to be 
destroyed by a subhalo. To account for this process in our model, we 
calculate, for the redshift interval of $z$ to $z + dz$, the minimum of 
pericentric radii $r_{\mt{min}}$ for subhalos' orbital motions, where 
their mass is larger than a supposed mass of a galactic disk 
($M_{\mt{th}}$). We then suppose that this minimum radius provides the 
upper limit for the radius of a galactic disk which can exist at $z$.
It is thus postulated that any galactic disk having a
size larger than this minimum of subhalos' pericentric radii will be 
destroyed by dynamical interaction with subhalos. In other words, the 
size of a galactic disk at $z$ is limited by the orbits of massive 
subhalos in the redshift interval of $z$ to $z + dz$.

We note that in the actual calculation of merger trees, there are only
a few subhalos with mass $M > M_{\mt{th}}$ from $z$ to $z + dz$, which makes 
it difficult to estimate the minimum of pericentric radii in the 
interval $z$ to $z + dz$ in a statistically significant manner. Instead, 
we first consider the cumulative distribution from $z$ to 
$z = 0$, for the minimum of pericentric radii for subhalos with $M > 
M_{\mt{th}}$. We then convert this cumulative form with respect to $z$ 
into the differential form from $z$ to $z + dz$. Notice that in this 
calculation we exclude the subhalos which end up with ``centrally 
merged''. Such subhalos sink to the center of the host halo, which 
results in major mergers with a galactic disk. Therefore, 
by excluding such subhalos, we can examine a dynamically quiet disk 
system that does not experience any major-merger events, thereby having 
a possibility to exist as a galactic disk.

\section{Results}\label{results}

\subsection{Distribution of Pericentric Radii of Subhalos}

\begin{figure}
\centerline{\includegraphics[width=9cm]{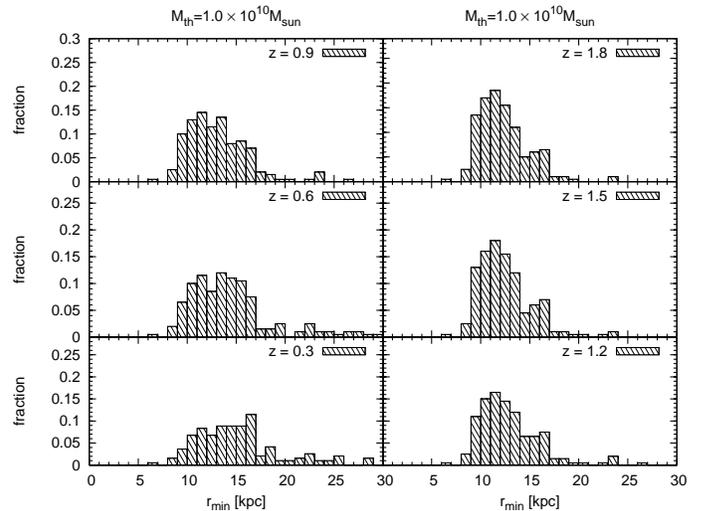}}
\caption{The histogram of 
the minimum of pericentric radii for orbiting subhalos with the mass 
larger than $M_{\mt{th}}=10^{10}M_\odot$ from the given redshift $z$ to 
$z = 0$ in each merger tree. In the upper right corner this $z$ 
is depicted. We exclude the subhalos which end up with centrally merged. 
The grid of the histogram is 1 kpc. All the panels are based on the 200 
merger trees.}
\label{fig-hist-1e10}
\end{figure}

\begin{figure}
\centerline{\includegraphics[width=9cm]{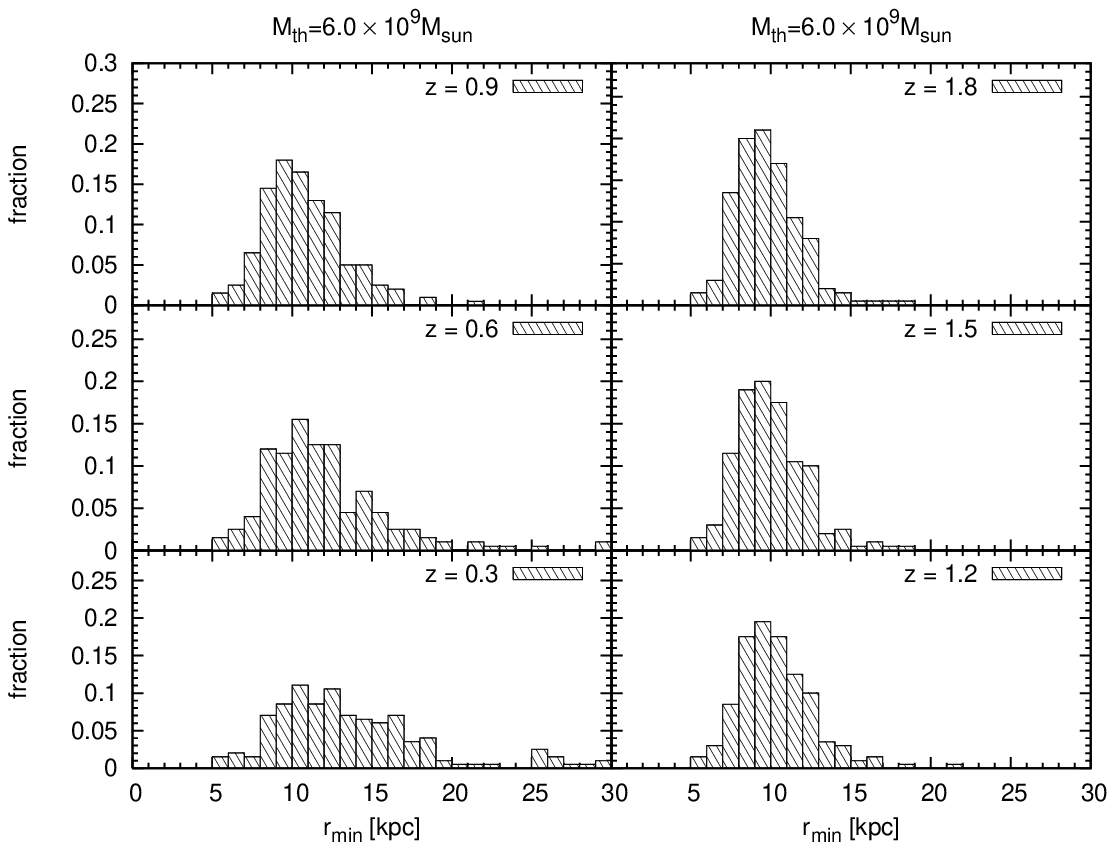}}
  \caption{The same as Figure \ref{fig-hist-1e10} but for
  $M_{\mt{th}}=6\times10^{9}M_\odot$.}
  \label{fig-hist-6e9}
\end{figure}

\begin{figure}
\centerline{\includegraphics[width=9cm]{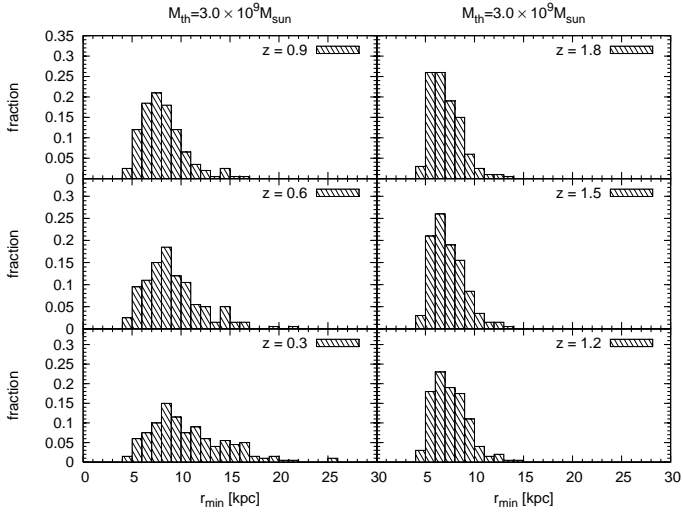}}
  \caption{The same as Figure \ref{fig-hist-1e10} but for
  $M_{\mt{th}}=3\times10^{9}M_\odot$.}
  \label{fig-hist-3e9}
\end{figure}

Here we show the fiducial case of a host halo with 
$M_{\mt{host}}=1.4 \times 10^{12}M_\odot$ at $z = 0$ and the minimum 
mass of merger tree is $M_{\mt{min}}=10^5M_\odot$. Figure 
\ref{fig-hist-1e10} - \ref{fig-hist-3e9} show the histogram of the 
minimum of pericentric radii for orbiting subhalos with $M > 
M_{\mt{th}}$ from $z$ to $z=0$. In these figures, we vary 
$z$ from 0.3 to 1.8. Figure \ref{fig-hist-1e10}, 
\ref{fig-hist-6e9}, and \ref{fig-hist-3e9} show, respectively, the cases 
of $M_{\mt{th}}=10^{10}M_\odot$, $6\times10^9M_\odot$, and 
$3\times10^9M_\odot$, based on 200 merger trees. It is found that
at fixed $M_{\mt{th}}$, $r_{\mt{min}}$ at 
higher $z$ is smaller than that at lower $z$. This can be 
explained as follows. At high $z$ the virial radius of a host halo 
is small as suggested from equation (\ref{virial-radius}). Also the 
orbital time of subhalos which accreted at higher $z$ is longer 
than those accreted at lower $z$. Thus the orbits of subhalos with 
longer orbital time decay more to inner radii since dynamical friction 
affects these subhalos during a longer time, thereby yielding smaller 
$r_{\mt{min}}$. On the other hand, at fixed $z$, $r_{\mt{min}}$ for 
larger $M_{\mt{th}}$ is larger than that for smaller $M_{\mt{th}}$. This 
may be because the accretion of massive subhalos occur at preferentially 
low $z$, compared with less massive ones.

\begin{figure}
\centerline{\includegraphics[width=9cm]{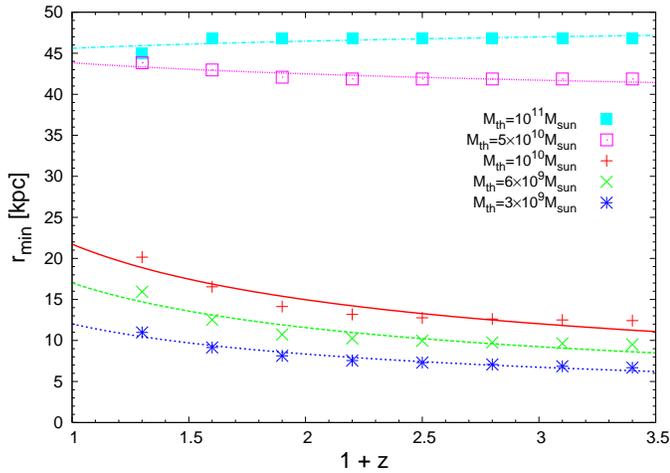}}
\caption{Ensemble mean of the minimum of pericentric radii for orbiting 
subhalos with the mass larger than $M_{\mt{th}}=10^{10}M_\odot$ (plus),
$6 \times 10^9 M_\odot$ (cross), and
$3 \times 10^9 M_\odot$ (asterisk) from redshift $z$ to $z = 0$.
Also, shown are the additional two cases with more massive subhalos,
$M_{\mt{th}}=10^{11}M_\odot$ (filled square) and $5 \times 10^{10} M_\odot$
(open square).
All the plots are based on the result of 200 merger trees.
All the lines are the results of the least-squares fitting to these plots
for the ensemble mean.
}\label{fig-hist-ave}
\end{figure}

Figure \ref{fig-hist-ave} shows the average of these histograms (plus,
cross, and asterisk).
The lines denote the fitting functions in terms of the least-squares method
(without employing a dispersion-weighted fitting), which are given as
$22 (1+z)^{-0.54}$ for $M_{\mt{th}}=10^{10}M_\odot$,
$17 (1+z)^{-0.56}$ for $M_{\mt{th}}=6\times10^9M_\odot$, and
$12 (1+z)^{-0.53}$ for $M_{\mt{th}}=3\times10^9M_\odot$. If a dispersion is used
so that a fitting is weighted in favor of a low dispersion, we obtain somewhat
shallower fitting curves: $19 (1+z)^{-0.38}$, $13 (1+z)^{-0.27}$, and
$11 (1+z)^{-0.42}$, respectively. These plots show clearly that
at fixed $M_{\mt{th}}$, $r_{\mt{min}}$ at high $z$ is smaller than that
at low $z$ and at fixed $z$, $r_{\mt{min}}$
with massive $M_{\mt{th}}$ is larger than that with less massive one.

In addition to these cases with a threshold subhalo mass of
$M_{\mt{th}} \le 10^{10}M_\odot$, we additionally consider much more massive cases,
i.e., $M_{\mt{th}}=5\times10^{10}M_\odot$ and $10^{11}M_\odot$ in view of
recent studies (Stewart et al. 2008; Purcell et al. 2009) showing that
such massive subhalos are most crucial in the survival of a thin disk.
In Figure \ref{fig-hist-ave}, open and filled squares show the cases of these
threshold masses, respectively. It is found that such massive subhalos,
which do not end up with the category of centrally merged ones, remain
at large orbital distances of $> 40$~kpc and that the distribution of
their pericentric distances stays nearly constant: the fitting functions are
given as $44 (1+z)^{-0.05}$ and $46 (1+z)^{+0.03}$ for
$M_{\mt{th}}=5\times10^{10}M_\odot$
and $10^{11}M_\odot$, respectively, for both with and without using
a dispersion in the fitting procedure. Thus, a disk which is exempt from
the central merging of such massive subhalos is likely able to have
a large radius, unless the effects of less massive ones are significant.

To derive the distribution from $z$ to $z + dz$, we convert these 
fitting functions into differential forms as 
follows. We define $N(z, R_{\mt{size}}, M_{\mt{disk}})$ as the 
distribution function of galactic disks with the size, $R_{\mt{size}}$, 
and mass, $M_{\mt{disk}}$, at $z$. Considering the average 
of the size of galactic disks from the redshift $z$ to $z = 0$, then we 
can write this as,
\begin{equation}
 \left<R_{\mt{size}}\right>=\frac{\int_0^\infty R'_{\mt{size}}\int_z^0  
N(z', R'_{\mt{size}}, M_{\mt{disk}}) 
dz' R'_{\mt{size}}}{\int_0^\infty\int_z^0 N(z',  
R'_{\mt{size}}, M_{\mt{disk}}) dz'dR'_{\mt{size}}}.
\end{equation}
Assuming that all disks are destroyed when subhalos with mass larger 
than the threshold mass $M_{\mt{th}}$ (comparable to a disk mass) interact with 
anywhere inside a galactic disk, $\left<R_{\mt{size}}\right>$ agrees 
with the minimum of pericentric radius for orbiting subhalos with mass 
larger than $M_{\mt{th}}$ from the redshift $z$ to $z = 0$. We define 
the distribution of galactic disks with mass smaller than 
$M_{\mt{disk}}$ from $z$ to $z + dz$ as $\overline{R_{\mt{size}}}$. We 
can write this as
\begin{equation}
 \overline{R_{\mt{size}}}=\frac{\int_0^\infty R'_{\mt{size}}N(z, 
R'_{\mt{size}}, M_{\mt{disk}})dR'_{\mt{size}}}{\int_0^\infty 
N(z, R'_{\mt{size}}, 
M_{\mt{disk}})dR'_{\mt{size}}}.
\end{equation}
With normalization $\int_0^\infty N(z, R'_{\mt{size}}, 
M_{\mt{disk}})dR'_{\mt{size}}=1$, we obtain
\begin{equation}
\left<R_{\mt{size}}\right>
= \frac{1}{z}\int_0^z dz' \overline{R_{\mt{size}}}(z').
\end{equation}
Assuming $\left<R_{\mt{size}}\right>=\alpha(1+z)^\beta$
and employing differentiation, we obtain
\begin{equation}
\overline{R_{\mt{size}}}(z) = \alpha(1+z)^{\beta-1}[1+(1+\beta)z].
\end{equation}
In this way, we can convert the cumulative function over $[0, z]$ into the 
differential function over $[z, z+ dz]$.

\begin{figure}
\includegraphics[]{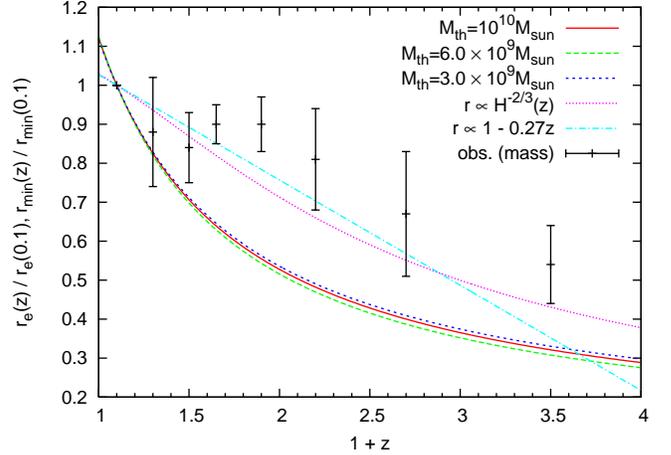}
\caption{
Redshift dependence of disk-size evolution in various disk models
and observed disks, where comparison is made by normalizing
a disk size to be 1 at $z = 0.1$. Solid, long-dahsed, and short-dashed
lines denote the fitting results (without considering
the dispersion) to the average of the minimum pericentric radius $r_{\mt{min}}$
in subhalos with mass which is larger than $M_{\mt{th}}=10^{10}M_\odot$,
$6\times 10^{9}M_\odot$, and $3\times 10^{9}M_\odot$, respectively,
from $z$ to $z+ dz$. Also shown are other model predictions:
dotted line is the expected evolution of disk scale length
from Mo et al. (1999) $r_e(z)/r_e(0.1)\propto H^{-2/3}(z)$,
and dash-dotted line is the expected evolution of
disk scale length from BS02 $r_e(z)/r_e(0.1)\propto1-0.27z$,
where $r_e$ is the effective radius. 
The points denote the evolution of an average disk 
scale length from observational result (Trujillo et al. 2006),
$r_e(z)/r_e(0.1)$.
}
  \label{fig-average-evolv-normalize}
\end{figure}

Figure \ref{fig-average-evolv-normalize} shows these differential 
functions normalized at $z = 0.1$, so that redshift dependence
of size evolution is highlighted. It is evident that there 
is no significant difference in redshift dependence between our models
with different mass thresholds $M_{\mt{th}}$. We also plot the prediction
of the existing disk evolution models: Mo, Mao \& White (1999) model of
$r_e(z)/r_e(0.1)\propto H^{-2/3}(z)$ (dotted line) and
BS02 model of $r_e(z)/r_e(0.1)\propto1-0.27z$ (dash-dotted line),
where $r_e$ is the effective radius. Also shown (points with error bars)
is the evolution of average disk scale length from observational results
(Trujillo et al. 2006). It follows that the redshift dependence of
all the characteristic sizes is basically similar to each other,
especially at $z \ga 1$.

\subsection{Comparison with Observed Disk Sizes}

Our interest in this work is to elucidate whether a galactic disk survives
in the CDM model. For this purpose, we need to estimate the likely outer
edge of an observed disk, which generally has an exponential light 
distribution. For a class of disk galaxies, this light distribution does not 
extend to an arbitrarily large radius, but is radially truncated at the 
outer part. This radius is a so-called truncation radius, which can be regarded
as the outer edge of the disk. Observations of nearby disk galaxies indicate
that this truncation radius is related to the disk scale length as
$R_{\mt{break}}\approx 2 R_d - 4 R_d$ (e.g. Kregel et al. 2002). In contrast,
at high $z$ few observations are available for the information of
a truncation radius. We simply assume the relation $R_{\mt{break}}= 3.5 R_d$
at current epoch for such remote disk galaxies.

\begin{figure}
\includegraphics[]{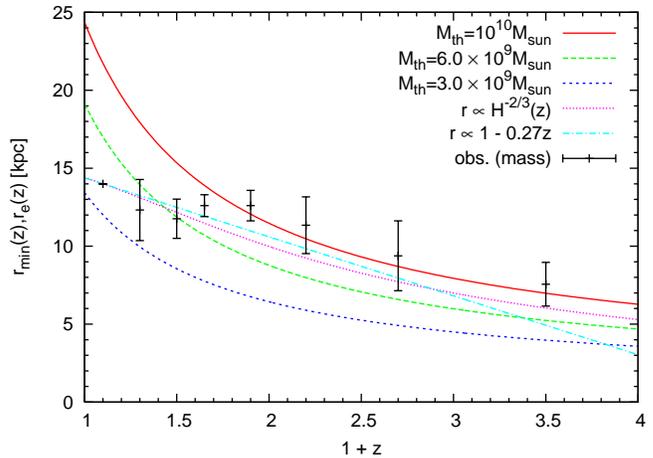}
\caption{The same as Figure \ref{fig-average-evolv-normalize} but without being 
normalized. We assume a Milky Way type disk, where 
$r_e(0.1)=4~\mathrm{kpc}$ and the size of a disk is $3.5r_e$.}
\label{fig-average-evolv}
\end{figure}

We consider the Milky Way-type disk with mass $10^{10}M_{\odot}$ and
the scale length $R_d=4~\mathrm{kpc}$ for the present experiment,
to obtain the evolution of the disk's outer edge.
In Figure \ref{fig-average-evolv}, the points show the redshift evolution of
a supposed disk size at fixed mass as derived from an observed disk scale length.
Notice that we assume $r_e$ is equal
to $R_d$, since the effective radius is related to the scale length by a simple 
function. It follows that at low $z$ the points are well below our
theoretical upper limits, thereby suggesting that observed galactic disks
at low $z$ can survive in the CDM model.
On the other hand, at high $z$ there is 
a possibility that some of galactic disks may be destroyed by 
dynamical effects of subhalos, yielding some slight disagreement with
observations. This possibility can be precluded by other effects occurred in
a disk. For instance, when a galactic disk contains the large amount of
cold interstellar gas, such gas which is unaffected by subhalos can 
reduce the thickness of a galactic disk. Also, at the truncation radius the 
mass density of a galactic disk are low enough that the effect of subhalos
on a disk structure is minor.

\section{Discussion and Concluding Remarks}\label{discussion}

We have investigated the dynamical effect of subhalos on the size evolution of 
a galactic disk, based on a semi-analytic method that includes approximated 
and empirical relations for the evolution of subhalos as obtained in
detailed numerical simulations of the CDM model. This is motivated by
our previous work (Hayashi \& Chiba 2006) that subhalos are able
to induce the dynamical heating and tidal destruction of a galactic disk,
thereby affecting the distribution of the size of a disk which does not suffer
from the interaction with subhalos; such a disk ought to be smaller than
the minimum of pericentric orbital radii of subhalos. We have found
that the upper limit for the size of a galactic disk with large mass is 
larger than that with small mass. Also, with fixed mass, the upper limit for 
the size of a galactic disk at low $z$ is larger than that at high $z$.

Stellar-population analyses of present-day galaxies have revealed
the so-called 'downsizing' evolution, where 
the stars in more massive galaxies tend to have formed earlier and over 
a shorter time-span (Thomas et al. 2005; MacArthur et al. 2004). On the 
other hand, in the CDM model halos are formed hierarchically in a
bottom-up manner, contrary to the downsizing evolution.
In view of this general downsizing evolution of galaxies,
our result that a galactic disk with larger size and mass is available 
only at lower $z$ may conflict with observations. However, 
it is noted that this evolution of a disk is limited by the pericentric
orbital radius of subhalos, i.e., a disk which has never been affected
by subhalos, whereas a galactic disk having initially a large radius
can indeed be dynamically heated and/or tidally destructed by subhalos.
Even after such a destruction incident of a thin stellar disk, 
residual gas or infalling fresh gas allows to form a new disk component.
Indeed according to recent numerical simulations, infalling subhalos do not fully
destroy a galactic disk, where 
some fraction of the thin disk component are survived (Kazantzidis et al. 
2008; Villalobos \& Helmi 2008). We thus regard that old stellar components 
in massive disk galaxies may be the relic of a first-existing galactic disk. 
In particular, recent numerical simulations have demonstrated that
a thick disk can be formed by the dynamical interaction between such a first-existing
disk and subhalos (Hayashi \& Chiba 2006;
Kazantzidis et al. 2008; Villalobos \& Helmi 2008).

Several models have been proposed for the evolution of a disk scale length
(e.g., MMW98; BS02). However, such models do not explicitly take into account
the finite extent of a galactic disk.
For instance, the MMW98 model assumes that stars form from gas instantaneously,
while maintaining an exponential density profile of a disk. Since an
exponential disk does not have an outer limit in its profile, this model 
cannot provide the size evolution of a galactic disk. On the other hand, 
the BS02 model adopts the present state of a disk galaxy as an initial
condition for its backward approach in time. This model also assumes
an exponential stellar disk at present, so it is unclear how the
disk size evolves; we need to take into account explicitly the size evolution of
a disk in its model for the formation of an exponential density profile.

As such a model, we consider a so-called viscous disk model
(Lin \& Pringle 1987; Yoshii \& Sommer-Larsen 1989).
This model shows that the exponential distribution 
of stars in a galactic disk is a natural product of angular momentum 
redistribution caused by viscosity. Yoshii \& Sommer-Larsen (1989) 
found that if the condition that star formation time scale being 
comparable to viscous time scale is maintained, then the final 
stellar distribution shows an exponential profile irrespective of the 
specific form of the initial gas profile. However, since stars are assumed to
form continuously as long as gas is still remained, where fresh gas is always
replenished by infalling gas, it is evident in this model that no outer limit
comes out in a galactic disk. We thus need to consider, for instance,
the star formation threshold in this model, to obtain the size evolution of
a galactic disk, where stars are formed when gas density is beyond some
threshold value. Such a threshold in gas density may be defined by
a so-called Toomre criterion for gravitational instability of an axisymmetric
disk and/or some external feedback such as a UV background radiation, although
an explicit account of these processes is yet to be explored in a viscous disk model.

It is worth noting that in our models presented here, a couple of
other effects, e.g., gas-dynamical response of a disk and tidal force of
a disk on subhalos are ignored. When we consider cold gas components
in a disk, the stronger binding energy of such a disk than that without
gas may be less affected by subhalos. Also, the tidal force of a disk
on subhalos may make the mass loss of subhalos more significant. Both
of these effects help the survival of a galactic disk.
If we assume that the orbital circularity of subhalos is more tangential
than that adopted here, the number of subhalos having nearly circular orbits
becomes larger. Then subhalos can be more easily survived at larger orbital
radii, so the size of a galactic disk limited by such subhalos may be
larger. This therefore suggests that it is essential to understand the realistic
dynamical properties of subhalos, based on the CDM model, in order to
set tighter constraints on the evolution of a disk galaxy like the Milky Way.

We conclude that subhalos tightly regulate the structure and evolution of 
a galactic disk. The effects of massive subhalos 
on a galactic disk are significant compared to less massive ones, so
it is important to consider the dynamics and properties of 
such subhalos. Since at high $z$ subhalos can sink to the central 
region of a host halo, the dynamical effects of subhalos on a galactic 
disk is more efficient at high $z$ than at low $z$. Then, a 
galactic disk at present might experience the interaction with subhalos 
one or more times during its evolutionary process, especially at high $z$. We 
postulate that a thick disk is a relic of a pre-existing galactic disk 
after these interaction events.

It is also important to construct the new formation model of a galactic disk
taking into account its size evolution and dynamical effects of subhalos, so that
one can directly compare the model results with observed disks at high $z$.
In this respect, it is useful to carry out high-resolution hydrodynamical
simulations and predict detailed properties of galactic disks at high $z$.
Furthermore, more detailed observations of extragalactic thick disks,
in particular their kinematics, are needed to find the evidence for
the interaction between subhalos and a galactic disk (e.g., Herrmann, Ciardullo,
\& Sigurdsson 2009).

\acknowledgments
The authors are grateful to the anonymous referee for useful comments that
helped improve the manuscript.
This work has been supported in part by a Grant-in-Aid for
Scientific Research (20340039) of the Ministry of Education, Culture,
Sports, Science and Technology in Japan.



\end{document}